\documentclass[letter]{emulateapj}
\usepackage{graphicx}
\usepackage{apjfonts}

\newcommand{\sj}{XTE J1701--462}
\newcommand{\ledd}{$L_{Edd}$}
\newcommand{\mdot}{$\dot{m}$}
\newcommand{\mdisk}{$\dot{m}_{d}$}
\newcommand{\mfilt}{$<$$\dot{m}_{d}$$>$}

\shorttitle{\sj}
\shortauthors{Jeroen Homan et al.}

\begin{document}

\title{{\it RXTE} observations of the first transient Z source \sj:
\\ shedding new light on mass accretion in luminous neutron star
X-ray binaries}

\author{
Jeroen Homan\altaffilmark{1},
Michiel van der Klis\altaffilmark{2},
Rudy Wijnands\altaffilmark{2},
Tomaso Belloni\altaffilmark{3},
Rob Fender\altaffilmark{6},
Marc Klein-Wolt\altaffilmark{2,4},
Piergiorgio Casella\altaffilmark{2},
Mariano M\'endez\altaffilmark{2,5},
Elena Gallo\altaffilmark{7,8},
Walter H.G.\ Lewin\altaffilmark{1},
Neil Gehrels\altaffilmark{4}
}

\altaffiltext{1}{MIT Kavli Institute for Astrophysics and Space
Research, 70 Vassar Street, Cambridge, MA 02139;
jeroen@space.mit.edu}

\altaffiltext{2}{Astronomical Institute `Anton Pannekoek', University
of Amsterdam, Kruislaan 403,
1098 SJ, Amsterdam, the Netherlands}

\altaffiltext{3}{INAF-Osservatorio Astronomico di Brera, Via E.
Bianchi 46, I-23807 Merate (LC), Italy)}

\altaffiltext{4}{NASA/Goddard Space Flight Center, Greenbelt, MD 20771}

\altaffiltext{5}{SRON, Netherlands Institute for Space Research,
Sorbonnelaan 2, 3584 CA, Utrecht, the Netherlands}

\altaffiltext{6}{School of Physics and Astronomy, University of
Southampton, Southampton, Hampshire, SO17 1BJ}

\altaffiltext{7}{Department of Physics, University of California,
Santa Barbara, CA 93106}

\altaffiltext{8}{{\it Chandra} Fellow}

\begin{abstract} We report on the first ten weeks of {\it Rossi X-ray
Timing Explorer} observations of the X-ray transient \sj. Large count
rate variations were accompanied by changes in the source's spectral
and variability properties. Based on comparisons with other sources
we conclude that \sj\ had all the characteristics of the neutron star
Z sources, i.e., the class of the brightest persistent neutron star
low-mass X-ray binaries. These include the typical Z-shaped tracks
traced out in X-ray color diagrams on time scales of hours and the
variability components detected in the power spectra, such as kHz
quasi-periodic oscillations (QPOs) and normal and horizontal branch
oscillations. \sj\ is the first transient Z source and provides
unique insights into mass accretion rate (\mdot) and luminosity
dependencies in neutron star X-ray binaries. As its overall
luminosity decreased, we observed that \sj\ switched between two
types of Z-source behavior never previously identified as clearly in
the same Z source, with most of the branches of the Z-track changing
their shape and/or orientation. We interpret this switch as an
extreme case of the more moderate longterm changes seen in the
persistent Z sources and suggest that they are the result of changes
in \mdot. We also suggest that the Cyg-like Z sources (Cyg X-2, GX
5-1, and GX 340+0) might be substantially more luminous ($>$50\%)
than the Sco-like Z sources (Sco X-1, GX 17+2, and GX 349+2).
Adopting a possible explanation for the behavior of kHz QPOs, which
involves a prompt as well as a filtered response to changes in \mdot,
we further propose that changes in \mdot\ can explain both movement
along the Z track and changes in the shape of the Z track itself. We
discuss some consequences of this and consider the possibility that
the branches of the Z will smoothly evolve into the branches observed
in the X-ray color diagrams of the less luminous atoll sources,
although not in a way that was previously suggested.

\end{abstract}


\section{Introduction}\label{sec:intro}

Based on their correlated X-ray spectral and rapid variability
properties, the brightest persistent neutron star low-mass X-ray
binaries (NSXBs) are often divided in two classes: the `atoll' and
`Z' sources \citep{hava1989}, named after the shape of the tracks
they trace out in X-ray color-color diagrams (CDs) and
hardness-intensity diagrams (HIDs). The Z sources form the brightest
of these two classes and are believed to accrete at near-Eddington
luminosities \citep[0.5--1.0 \ledd,][]{va2006}.  

\begin{figure}[t]
\centerline{\includegraphics[height=8.5cm,angle=-90]{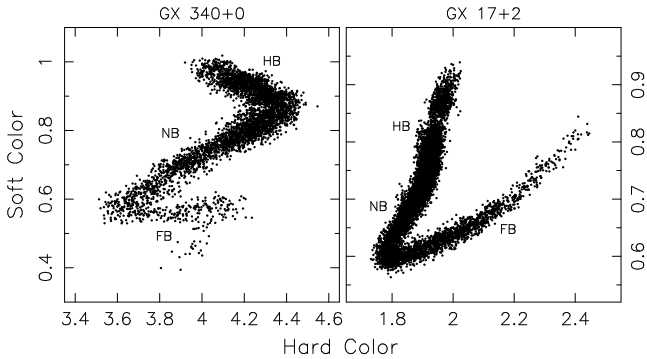}}
\caption{Color-color diagrams of the Z sources GX 340+0 (Cyg-like)
and GX 17+2 (Sco-like) from {\it RXTE}/PCA data. The diagrams were
constructed in a similar way as described in \S\ref{sec:analysis}.
Each data point represent 32s of data. The typical Z source branches
have been labeled, with HB, NB, and FB (see \S\ref{sec:intro} for
more details).} \label{fig:z_source} \end{figure}

Six Galactic NSXBs have been classified as a Z source: Sco X-1, GX
17+2, GX 349+2, GX 5-1, GX 340+0, and Cyg X-2. They are characterized
by three-branched tracks in their CDs (see Figure \ref{fig:z_source}
for two examples), and HIDs, which in some cases resemble the
character `Z'. In other cases they have a more `$\nu$'-like shape,
but we will refer to them as Z tracks nevertheless. From top to
bottom, the three branches of the Z track are referred to as the
horizontal branch (HB), the normal branch (NB), and flaring branch
(FB). Only in Cyg X-2, GX 340+0, GX 5-1 \citep[the so-called
`Cyg-like' Z sources,][]{kuvaoo1994} is the HB nearly horizontal,
particularly in the HID. In GX 17+2 and Sco X-1, the HB is nearly
vertical, and in GX 349+2 no full-fledged HB has been observed so
far. On the other hand, only in the latter three sources (the
`Sco-like' sources) the FB can be identified with strong flaring,
with count rates significantly higher than observed on the other two
branches, whereas the Cyg-like sources sometimes show count rate
decreases on that branch. 

Despite these (and various other) differences among the Z sources, it
has often been assumed that in all of them the mass accretion rate
(\mdot) increases monotonically from the HB, via the NB, to the FB.
There are several observational results that seem to support this:
(1) the typical frequencies of most variability components (which in
many models scale with \mdot) increase from the HB to NB, (2) the
optical and UV flux, often thought to be the result of reprocessed
X-rays, increase from the HB to FB \citep[see,
e.g.,][]{havaeb1990,vrraga1990}, (3) jumps between branches are not
observed, in accordance with the assumption that \mdot\ changes
continuously and not abruptly. However, this interpretation implies
that X-ray luminosity does not track \mdot.

Motion along the branches of the Z usually takes place on time scales
of hours to days. On longer time scales it has been observed that the
entire Z track changes its location in the CD and HID, or even
changes its morphology, as is best observed in Cyg X-2
\citep{kuvava1996,wivaku1997}. These longterm changes (or `secular
changes' as they are often referred to) appear to be strongest in the
Cyg-like sources and it has been suggested that this is because the
Cyg-like sources might have a higher inclination
\citep{kuvaoo1994,kuva1995}. More recently, \citet{hovajo2002}
suggested that the longterm changes of the Z tracks might be the
result of changes in \mdot, which would be in apparent conflict with
the view that \mdot\ changes along the Z track.  However, based on
the observation that otherwise similar spectral/timing states can
occur at different luminosity levels both across sources and in a
given source, \citet{va2001} proposed a specific alternative
quantitative measure to determine source state, namely \mdot\ through
the disk, \mdisk, normalized by its own longterm average. Changes in
\mdisk\ could then perhaps be responsible for motion along the Z
track {\it and} changes in the Z track itself (see \S4.2).

The power spectra of the Z sources show several types of
quasi-periodic oscillations (QPOs) and noise components \citep[for
reviews, see][]{va1995a,va2006}.  Their presence and properties are
strongly correlated with the position of the source along the Z track
\citep{hava1989}.  Three types of low frequency ($<$100 Hz) QPOs are
seen in the Z sources: the horizontal branch (HBOs), normal branch
(NBOs) and flaring branch oscillations (FBOs). Their names derive
from the branches on which they were originally found. The HBO is
found on the HB and NB with a frequency (15--60 Hz) that gradually
increases along the HB towards the NB. The NBO and FBO are most
likely part of one phenomenon. They are found on the NB and FB (near
the NB/FB vertex) but not on the HB. On the NB the QPO has a
frequency of $\sim$5--7 Hz, which (only in the Sco-like sources)
rapidly increases to $\sim$20 Hz when the source moves across the
NB/FB vertex \citep{prhale1986,diva2000,cabest2006}. Twin  kHz QPOs
have been found in all Z sources. They are often observed
simultaneously, with a frequency difference of $\sim$300 Hz, and have
frequencies between 215 Hz and 1130 Hz, which increase from the HB to
the NB \citep[in Sco X-1 also onto the FB,][]{vaswzh1996}. Two types
of noise are seen in the Z sources. These are the very low frequency
noise (VLFN) and the low frequency noise (LFN). The VLFN, which is
found at frequencies below 1 Hz, can be described by a power law and
is found on all branches, whereas the LFN, which is peaked or
flat-topped noise component (with cutoff frequencies of 2--10 Hz), is
only observed on the HB and NB.

In addition to the six sources that were classified as Z sources by
\citet{hava1989} a few other persistent sources have shown properties
that are characteristic of the Z sources. In its high luminosity
state Cir X-1 has shown tracks in its HID that were identified with
the HB, NB, and FB by \citet{shbrle1999}. These authors also detected
QPOs on the HB and NB that evolved in similar ways as the HBO and NBO
in the Z sources. \citet{bowiva2006} recently also reported the
discovery of twin kHz QPOs in Cir X-1. At lower luminosities,
however, Cir X-1 is similar to the atoll sources \citep{oovaku1995}.
Two other sources, both extra-galactic, were tentatively classified
as Z sources based on their tracks in a CD: LMC X-2 \citep[changes in
VLFN consistent with Z sources]{smhoku2003} and RX J0042.6+4115 in
M31 \citep[ a very incomplete Z track was observed and no timing
information was available, so classification is
uncertain]{bakoos2003}.

The atoll sources distinguish themselves from the Z sources not only
by their lower luminosities ($\sim$0.01--0.5 $L_{Edd}$) and different
tracks in CDs, but in general they also have harder spectra and show
stronger variability than the Z sources. \citet{murech2001} and
\citet{gido2002} proposed to identify certain structures that atoll
sources display in the CD, in luminosity ranges well below those of Z
sources,  with HB, NB and FB. The rapid X-ray variability, and the
order in, and timescales on, which sources trace out the branches are
however not in accordance with an identification of this type
\citep{baol2002,vavame2003,revava2004,va2006}. Most of the
differences between the Z and atoll sources are likely the result of
a lower \mdot\ in the latter group. However, it is still unclear if
\mdot\ alone can explain all the observed differences. In particular
the four most luminous atoll sources GX 3+1, GX 9+1, GX 9+9 and GX
13+1 appear to have luminosities close to and perhaps overlapping
those of Z sources but quite different color and timing properties.
Early on it was argued that a difference in strength of the neutron
star magnetic field, possibly as the result of different evolutionary
history, played a role in the observational appearance of the two
classes \citep{hava1989}; the arguments for this were tied to
specific models for the rapid variability
\citep{alsh1985,lashal1985}  and the spectra \citep{pslami1995} of
these sources. However, the discovery of kHz QPOs and millisecond
X-ray pulsars have cast doubts on these models, and for certain QPO
models the similarity in the range of kHz QPO frequencies would
require a fine-tuned balance between magnetic field and \mdot\
\citep{whzh1997,fovame2000}.

The existence of Cir X-1, which can apparently switch between atoll
and Z-like behavior, suggests that \mdot\ alone might be enough to
explain all differences between the two types of sources. However,
even when most similar to a Z source Cir X-1 is still peculiar, and
most of the time the source displays behavior that is neither
characteristically Z nor atoll. Luminous transient NSXBs would be
ideal sources to further test the relation of source types to \mdot,
since they cover a large range in \mdot. However, while many NSXB
transients have shown atoll source characteristics, none has ever
been observed to show the full range of properties associated with Z
sources, likely because most of them have peak luminosities well
below \ledd. Until recently, the source that has come closest to
resembling a Z source is XTE J1806--246, a transient which during one
observation near the peak of its 1998 outburst
($\sim1.5\times10^{38}$ erg/s for 8 kpc, 2--30 keV) showed an
NBO-like 7--14 Hz QPO that was found in a structure in the CD that
resembled the Z source NB/FB \citep{wiva1999b}. The source was poorly
sampled near the peak of the outburst, so additional Z-like
properties may have been missed. At lower luminosities XTE J1806--246
displayed atoll-like behavior.

\subsection{\sj}

\sj\ was first detected with the All-Sky Monitor (ASM) on-board the {\it
Rossi X-ray Timing Explorer} ({\it RXTE}) on 2006 January 18
\citep{reli2006}. A reanalysis of the ASM data puts the start of the
outburst between 2005 December 27 and 2006 January 4 (Ron Remillard,
private communication). Initial observations with the Proportional
Counter Array (PCA) on-board {\it RXTE}  revealed QPOs near 6 Hz and
55 Hz and large luminosity swings on time scales of a few minutes
\citep{stswmo2006}, but did not lead to a  classification of the
source. Follow-up observations \citep{hobeva2006b,hobeva2006a},
however, suggested that \sj\ is the first new Z source in nearly 35
years, with the source exhibiting the typical Z tracks in the CD, and
showing the evolution of the timing properties along these tracks
characteristic of Z sources. Optical/near-IR
\citep{mabane2006,maba2006} and radio counterparts \citep{fesada2006}
were suggested for \sj, which were later confirmed by a precise
Chandra localization of the source \citep{krjuch2006}. The source has
also been detected with {\it Swift} \citep{kemast2006} and {\it
INTEGRAL} \citep{prbasc2006}. In this paper we present an analysis of
the first ten weeks of {\it RXTE}/PCA observations of the outburst,
which at the time of writing is still ongoing.  The main goal of this
work is to show those results that firmly establish \sj\ as a Z
source, rather than present a complete analysis of the source. An
analysis of (quasi-)simultaneous {\it RXTE} and ATCA radio
observations will be presented by Fender et al. (in prep.).

\begin{figure}[t]
\centerline{\includegraphics[height=8.5cm,angle=-90]{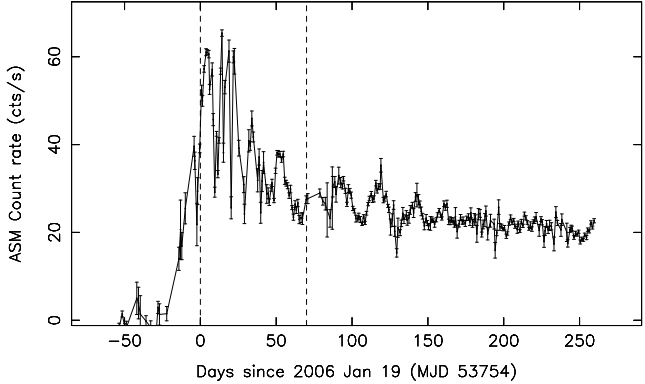}}
\caption{One-day averaged {\it RXTE}/ASM light curve of \sj. The two
dashed lines indicate the time interval studied in more detail with
{\it RXTE}/PCA  (see figure \ref{fig:curves}).} \label{fig:asm}
\end{figure}

\section{Observations and data analysis}\label{sec:analysis}

We have analyzed all 113 {\it RXTE}/PCA
\citep{brrosw1993,jamara2006} observations of \sj\ made
between 2006 January 19 and 2006 March 29 (MJD 53754--53824). Four
scanning observations, made on 2006 January 20 to improve the
positional accuracy, were excluded from our analysis, since they did
not yield useful data.

Background subtracted light curves with a time resolution of 16
seconds were constructed from the `Standard 2'-mode data, covering
the full effective PCA energy range ($\sim$2--60 keV, with 129
channels [1--129]), using only data from proportional counter unit
(PCU) number 2 (the most reliable of the five PCUs). No dead-time
corrections were applied, since these were relatively small: less
than 5.5\% for the highest count rates in our data set. We also
defined two X-ray colors, a soft color (SC) and a hard color (HC), as
the ratio of count rates (extracted from `Standard 2' data) in  the
$\sim$4.0--7.3 keV (channels 7--14) and $\sim$2.4--4.0  keV  (ch.\
3--6) bands (SC), and the  $\sim$9.8--18.2 keV (ch.\ 21--40) and
$\sim$7.3--9.8 keV (ch.\ 15--20) bands (HC). These colors were used
to produce color curves, CDs, HIDs, and soft color vs.\ intensity
diagrams (SIDs), with the intensity defined as the count rate in the
full PCA band (ch.\ 1--129).

\begin{figure}[t]
\centerline{\includegraphics[angle=-90,width=8cm]{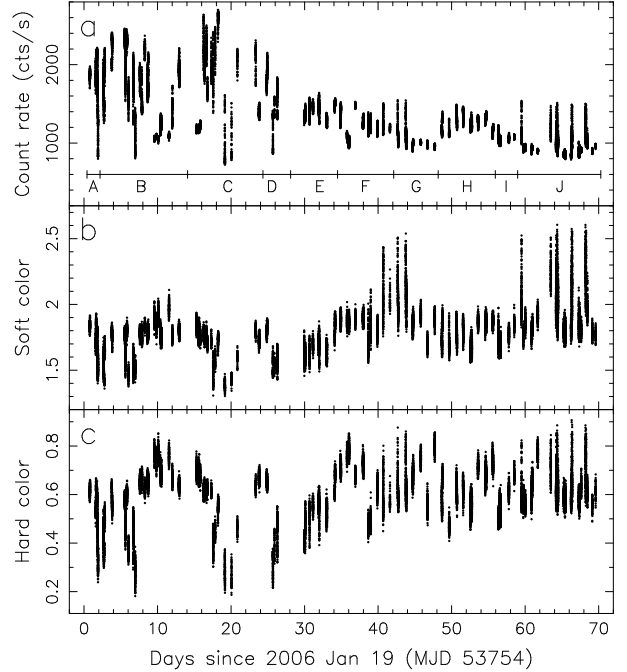}}
\caption{{\it RXTE}/PCA (a) light curve, (b) soft color curve, and
(c) hard color curve of \sj\ from the first ten weeks of our
observations. The time resolution is 16s. Only counts from PCU2 were
used. For the light curve we used its full energy range. For
definitions of colors see \S \ref{sec:analysis}. The characters A--J
in panel (a) indicate the ten intervals defined in \S
\ref{sec:results_curves}. See Figure \ref{fig:ccd} for corresponding
color-color diagrams.}\label{fig:curves} \end{figure}

\begin{figure*}[!t]
\centerline{\includegraphics[angle=-90,width=17cm]{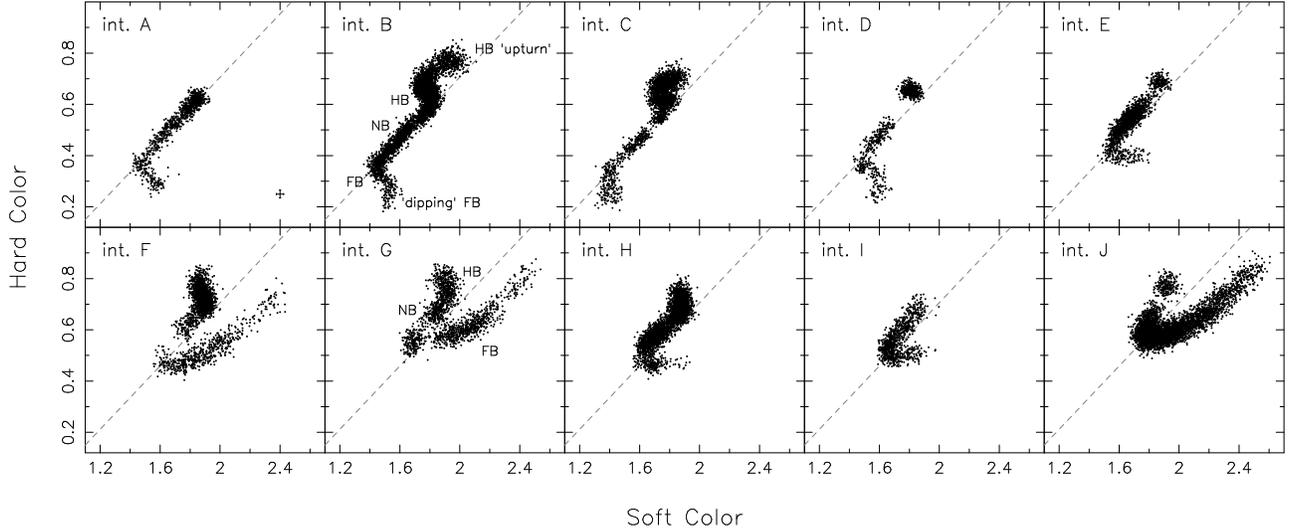}}
\caption{ Color-color diagrams for each of the ten intervals defined
in \S \ref{sec:results_curves}. Each dot represent 16s of data. See
Figure \ref{fig:curves}a for the corresponding time intervals. In the
panels for intervals B and G we have labeled the various branches of
the Z tracks. Representative errors bars are shown in the lower-right
of the interval A panel. The grey dashed diagonal lines can be used
to track changes in the location of the Z tracks, in particular the
NB/FB vertex, which moved along this line.}\label{fig:ccd}
\end{figure*}

\begin{figure}[!b]
\vspace{0.2cm}
\centerline{\includegraphics[width=8cm,angle=-90]{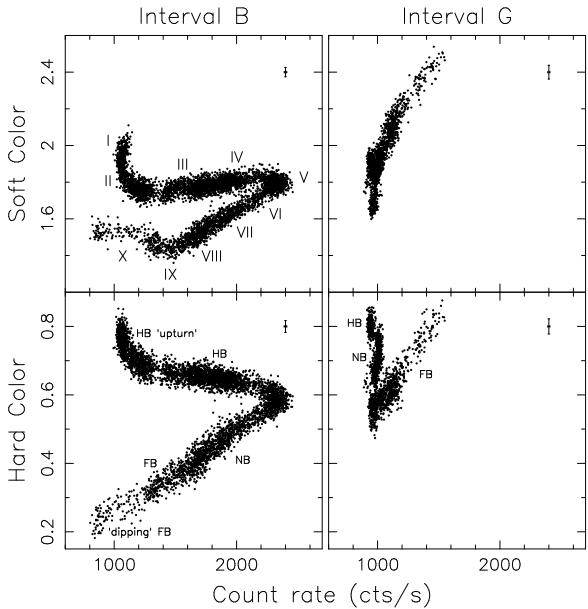}}
\caption{Two examples of soft color vs.\ intensity  and hard color
vs.\ intensity diagrams. Each dot represent 16s of data. The labels
I--X indicate the locations in the SID of interval B for which we
analyzed the associated power spectra, which are shown in Figure
\ref{fig:pds}. Labels for the different branches are shown in the
lower two panels and representative error bars are shown in the upper
right corners of each panel.} \label{fig:scd_hcd} \end{figure}

Power density spectra were created from high time resolution data
from all active PCUs summed together, using standard Fast Fourier
Transform (FFT) techniques \citep{va1989,va1995a}. The data were not
background corrected and no dead-time corrections were applied prior
to the FFTs; the effects of dead time were accounted for by our fit
function. We made power spectra from two energy bands, $\sim$2--60
keV and $\sim$6.9--60 keV, with lengths of 16 seconds and Nyquist
frequencies of 8192 Hz. The resulting power spectra were averaged,
based on various properties (e.g.\ time interval, count rate, color),
and rms normalized \citep{beha1990,mikiki1991,va1995b}.  These
normalized power spectra were fitted with a combination of several
Lorentzians ($P(\nu)=(r^2\Delta/\pi)[\Delta^2 +
(\nu-\nu_0)^2]^{-1}$), a constant (to represent the dead-time
modified Poisson level), and usually an additional power law
($P(\nu)\propto\nu^{-\alpha}$). Here $\nu_0$ is the centroid
frequency, $\Delta$ the half-width-at-half-maximum, and $r$ the
integrated fractional rms (from $-\infty$ to $\infty$). Instead of
$\nu_0$ and $\Delta$ we will quote the frequency $\nu_{max}$ at which
the Lorentzian attains its maximum in $\nu P(\nu)$ and the quality
factor, Q, where $\nu_{max}=\nu_0 (1 + 1/4Q^2)^{1/2}$ and
$Q=\nu_0/2\Delta$ \citep{bepsva2002}. The fractional rms amplitudes
quoted in this paper were the integrated power between 0 and $\infty$
for the Lorentzians, and between 0.1 Hz and 100 Hz for the power law
component. Errors on fit parameters were determined using
$\Delta\chi^2=1$. The low (below a few hundred Hz) and high frequency
parts of the power spectrum were usually fitted separately.

\section{Results and interpretation}\label{sec:results}

\subsection{Light curves and color evolution}\label{sec:results_curves}

In Figure \ref{fig:asm} we show the {\it RXTE}/ASM light curve of the
ongoing outburst of \sj. The source showed large flux variations
during the first $\sim$70 days of its outburst, which were followed
by a long period in which the source flux appeared to be slowly
declining. In this paper we focus on the {\it RXTE}/PCA data from the
time interval marked by the two vertical dashed lines in Figure
\ref{fig:asm}. Figure \ref{fig:curves} shows the {\it RXTE}/PCA light
curve and color curves of \sj\ from this interval. Immediately clear
from the light curve are the strong variations in the 2--60 keV count
rate, by factors up to 2, on time scales of hours to days. These
large count rate changes were accompanied by large color changes.
Initially we created single CDs, HIDs, and SIDs of all observations
combined, but these diagrams revealed that on a time scale of days to
weeks the shape of the tracks traced out by \sj\ changed
considerably. For this reason we decided to divide our data set into
different intervals; starting from the first observation we defined a
new interval every time when in a new observation the pattern traced
out in the CD had clearly shifted compared to the previous
observations. This method led to 10 intervals, labeled A to J, whose
corresponding times are indicated in Figure \ref{fig:curves}a. The
CDs of each interval are shown in Figure \ref{fig:ccd}. Small shifts
in count rate remained present in the HIDs/SIDs with the current
choice of intervals. However, these shifts did not show up in the
CDs.

The CDs of the 10 intervals shown in Figure \ref{fig:ccd} reveal a
large variety in shapes. Many of those shapes resemble the tracks
traced out by the Z sources and most of them are either similar to
those of the Cyg-like Z sources GX 340+4 and GX 5-1 \citep[intervals
A-D - see, e.g.,][for comparison]{jowiva1998,jovaho2002} or the
Sco-like Z sources \citep[intervals F, G, and J - see, e.g.,][for
comparison]{hovajo2002,onkuso2002,brgefo2003}. Based on these
similarities (compare also with Figure \ref{fig:z_source}) we can
identify the branches of the tracks in Figure \ref{fig:ccd} with
those seen in the Z sources (HB, NB, and FB), although support from
the power spectra is required to confirm these identifications.
Branch labels have been plotted in Figure \ref{fig:ccd} for the
tracks of intervals B and G. The NB/FB vertex of the tracks seemed to
move along a straight line in the CD, as the shape of the Z track
changed (see Fig. \ref{fig:ccd}).

\begin{figure*}[t]
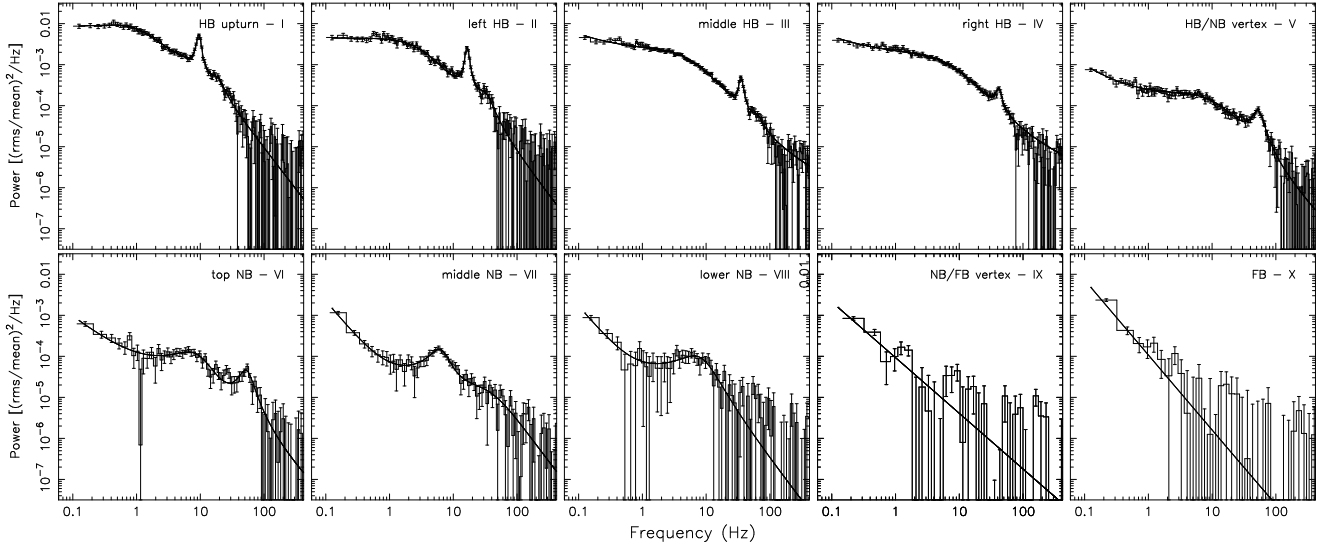

\centerline{\hbox{
\includegraphics[height=3.3cm]{f6a.eps}
\includegraphics[height=3.3cm]{f6b.eps}
\includegraphics[height=3.3cm]{f6c.eps}
\includegraphics[height=3.3cm]{f6d.eps}
\includegraphics[height=3.3cm]{f6e.eps}
}}
\centerline{\hbox{
\includegraphics[height=3.86381cm]{f6f.eps}
\includegraphics[height=3.86381cm]{f6g.eps}
\includegraphics[height=3.86381cm]{f6h.eps}
\includegraphics[height=3.86381cm]{f6i.eps}
\includegraphics[height=3.86381cm]{f6j.eps}
}}
\caption{Ten
power spectra (6.9--60 keV) from interval B. All power spectra are
rms normalized and the Poisson level has been subtracted. The solid
line in each panel shows the best fit to the power spectrum. See
Figure \ref{fig:scd_hcd} for the corresponding locations in the SID
of each power spectrum.}\label{fig:pds} \end{figure*}

The most striking difference between the CDs of the Cyg-like an
Sco-like intervals was the appearance of the FB.  In the Cyg-like
intervals the FB pointed downward - it was nearly vertical in
interval C - and was relatively short. In the Sco-like intervals the
FB pointed upwards and was much longer. Intervals E, H, and I seemed
to be intermediate between the Cyg-like and Sco-like intervals; their
FBs were short, like in the Cyg-like intervals, but were  nearly
horizontal. 
Other differences between the intervals could be seen in the behavior
of the HB, which was only clearly observed in intervals B, C, and
F--H. In intervals B and C the HB pointed to the upper left and
showed a prominent upturn to the upper right. In intervals F--H such
an upturn was absent and the HB had become nearly vertical. The
orientation of the NB in the CDs seemed to remain rather constant,
but its length appeared to be shorter in the Sco-like intervals. As
will be discussed in \S\ref{sec:compare}, examples of these types
of behavior have all been previously seen in other Z sources.

Changes in the shape and position of the Z tracks were also seen in
the SIDs and HIDs. In Figure \ref{fig:scd_hcd} we show example SIDs
and HIDs for intervals B and G. These two intervals showed nearly
complete tracks in their CDs and represent two opposites in terms of
observed behavior in the SID and HID.  The HB and NB went from
showing large count rate changes in intervals B and C to showing
nearly constant count rates in intervals G and J. Interval F, which
appeared Sco-like in the CD, still showed considerable changes in
count rate on the HB, but nearly constant rates on the NB. The
`intermediate' interval H showed moderate variations in the count
rate on the HB and NB. Large changes were also observed in the
properties of the FB: in intervals A--D the count rates decreased
from the HB/NB vertex all the way to the end of the FB. In interval E
complex behavior was observed, with decreases on the lower FB and
increases on the upper FB; the FB in interval F was similar, but the
lower FB showed an additional loop, back through the NB, similar to
what has been seen in Cyg X-2 \citep{kuvava1996,wivaku1997}.
Intervals G and J, on the other hand, showed count rate increases
from the NB/FB vertex to the top of the FB. Note that the highest
count rates in interval B were reached at the HB/NB vertex, whereas
in interval G the count rates peaked at the top of the FB. In
interval B, the top of the FB actually corresponded to the {\it
lowest} count rates. Despite minor differences, which will be
discussed in \S\ref{sec:compare}, most of the behavior seen in the
HIDs of \sj\ has also been observed in the other Z sources.

\subsection{Rapid X-ray variability} \label{sec:results_variability}

Given the similarities with the Z sources in the X-ray color
diagrams, the results of our timing analysis of \sj\ will be
discussed in terms of variability components commonly
observed in the Z sources.

\subsubsection{Broad band power spectra}

Like in other Z sources we found that the power spectra of \sj\
changed considerably as the source moved along its Z track. Since
interval B had the most complete track in the CD of all intervals, we
chose to use that interval to study the evolution of the broadband
power spectra in more detail. Similar evolution as described below
was observed in the other intervals. However, since our main aim is
to establish \sj\ as a bona fide Z source, a full analysis of all the
other intervals is beyond the scope of this paper. All values quoted
for the noise and QPO properties were measured in the 6.9--60 keV
band.

In Figure \ref{fig:scd_hcd} we have marked ten locations along the
track in the SID of interval B for which we have extracted power
spectra. The corresponding power spectra, which we will refer to,
using Roman numerals, as PDS I-X, together with our best fits, are
shown in Figure \ref{fig:pds}. Best fit parameters are given in Table
\ref{tab:fits}. The fits had reduced $\chi^2$ values between 0.81 and
1.25. Figure \ref{fig:pds} reveals clear changes in the shape of the
PDS and a decrease in the overall strength of the 1/16--100 Hz
variability (see also Table \ref{fig:pds}). On the HB (PDS I--IV),
where the power spectrum was dominated by a peaked/flat-topped noise
component, the total strength of the 1/16--100 Hz variability
smoothly decreased from $\sim$22\% rms to $\sim$16\% rms. A large
drop in the strength of the variability occurred near the HB/NB
vertex (PDS V) where we measured $\sim$7.7\% rms. At that point a
power-law noise component became clearly visible in the power
spectrum. On the NB (PDS VI--VIII) the variability gradually weakened
to $\sim$3.5\% rms near the NB/FB vertex (IX), and on the FB (PDS X),
where only power-law noise was detected, the strength of the
variability was 4.3$\pm$1.4\% rms.

\begin{table*} \caption{Broad-band power and best fit parameters of
the ten power spectra from interval B}\label{tab:fits}

\small{

\begin{tabular}{ccccccccccccc}

\hline
\hline
\multicolumn{1}{c}{} & \multicolumn{1}{c}{} & \multicolumn{2}{c}{VLFN} & \multicolumn{1}{c}{} & \multicolumn{3}{c}{LFN/NBO} & \multicolumn{1}{c}{} & \multicolumn{3}{c}{HBO} \\
\cline{3-4} \cline{6-8} \cline{10-12}
PDS & rms$^a$  & rms & PLI & & rms  & Q & $\nu_{max}$ & & rms  & Q & $\nu_{max}$  \\
    & (\%) &(\%) &     & & (\%) &   &  (Hz)       & & (\%) &   & (Hz)         \\
 
\hline
I	& 21.9$\pm$0.7	& \ldots	         &               & & 13.1$\pm$0.7	  & 0$^c$	      & 6.6$\pm$1.7	 &  & 11.0$\pm$0.2	   & 5.8$\pm$0.3	 & 9.55$\pm$0.03  \\
	&		&                        &               & & 12.5$\pm$0.9	  & 0.18$\pm$0.07     & 1.17$\pm$0.13	 &  &			   &			 &		  \\
II	& 19.2$\pm$0.5	& \ldots                 &               & & 14.8$^{+0.1}_{-0.6}$ & 0$^c$	      & 3.1$\pm$0.2	 &  & 10.2$^{+0.2}_{-0.8}$ & 6.0$^{+1.0}_{-0.3}$ & 16.61$\pm$0.05 \\
III	& 17.1$\pm$0.4	& 6.4$\pm$0.7            & 0.86$\pm$0.06 & & 14.51$\pm$0.19	  & 0$^c$	      & 6.04$\pm$0.13	 &  & 6.07$\pm$0.14	   & 6.1$\pm$0.4	 & 35.55$\pm$0.11 \\
IV	& 16.2$\pm$0.4	& 7.6$\pm$0.6            & 0.76$\pm$0.06 & & 13.9$\pm$0.2	  & 0$^c$	      & 7.5$\pm$0.2	 &  & 5.2$\pm$0.3	   & 4.2$\pm$0.6	 & 42.0$\pm$0.4   \\
V	& 7.77$\pm$0.14	& 2.0$^{+0.6}_{-0.4}$    & 1.10$\pm$0.17 & & 6.0$\pm$0.3	  & 0.08$\pm$0.07     & 11.5$\pm$0.8	 &  & 4.78$\pm$0.18	   & 2.2$\pm$0.2	 & 53.6$\pm$0.6   \\
VI	& 6.28$\pm$0.19	& 1.9$^{+0.6}_{-0.3}$    & 1.18$\pm$0.20 & & 4.2$\pm$0.3	  & 0.53$\pm$0.14     & 9.1$\pm$0.8	 &  & 4.3$\pm$0.4	   & 1.8$\pm$0.5	 & 53.6$\pm$1.7   \\
VII	& 5.0$\pm$0.4	& 1.61$\pm$0.09          & 1.84$\pm$0.15 & & 3.0$\pm$0.3$^b$	  & 1.22$\pm$0.25$^b$ & 6.5$\pm$0.3$^b$  &  & \ldots		   & \ldots		 & \ldots	  \\
        &               &                        &               & & 3.6$\pm$0.5	  & 0.2$\pm$0.4       & 39$\pm$11	 &  &			   &			 &		  \\
VIII	& 4.8$\pm$0.5	& 1.41$\pm$0.16          & 1.82$\pm$0.26 & & 3.7$\pm$0.2	  & 0.60$\pm$0.15     & 8.4$\pm$0.8	 &  & \ldots		   & \ldots		 & \ldots	  \\
IX	& 3.5$\pm$0.7	& 2.32$^{+0.34}_{-0.19}$ & 1.38$\pm$0.13 & & \ldots               & \ldots            & \ldots           &  & \ldots		   & \ldots		 & \ldots	  \\   
X	& 4.3$\pm$1.4	& 2.95$\pm$0.13          & 1.83$\pm$0.13 & & \ldots               & \ldots            & \ldots           &  & \ldots		   & \ldots		 & \ldots	  \\
\hline\\
\end{tabular}
}

$^a$ broad-band rms in the 1/16--100 Hz range\\
$^b$ NBO parameters\\
$^c$ parameter was fixed
\end{table*}

The peaked (or sometimes flat-topped) noise component that dominated
on the HB was fitted with a Lorentzian (two were needed in PDS I).
The Q-value of these components was either fixed at 0 or less than
0.6 (when not fixed). As the source moved along the HB onto the NB
(PDS VI), the characteristic frequency $\nu_{max}$ of this component
increased from $\sim$3 Hz to  $\sim$10 Hz. It is clear from Figure
\ref{fig:pds} that the peaked noise became weaker in the direction of
the NB, decreasing from $\sim$18\% rms in PDS I to $\sim$4\% in PDS
VI. At some point near the middle of the NB (PDS VII) the peaked
noise was replaced by, or evolved into, a broad QPO, which we
identified as the Z source normal branch QPO (NBO); the NBO is
discussed in more detail below. However, in addition to the NBO we
detected a noise component with a frequency of ~40 Hz, which is
probably not related to the peaked noise on the upper NB, but might
be a (broad) remnant of the HBO. On the lower NB (PDS VIII) we
detected a broad component with properties similar to the peaked
noise in PDS VI. No peaked or flat-topped noise was detected at the
NB/FB vertex and on the FB, including the dipping part (see PDS IX
and X).

The power-law noise component emerged near the middle of the HB. As
the source moved along the HB to the lower NB, this component, which
was fitted with a single power law, decreased in strength from 
$\sim$7\% rms (0.1--100 Hz) to 1.4\% rms. On the FB its strength
increased again to $\sim$4\% rms. Moving from the HB to the FB the
power-law index increased from $\sim$0.8 to $\sim$1.8, with only PDS
IX departing from this trend (with an index of 1.35$\pm$0.12).

Based on comparisons with other Z sources, the peaked and power-law
noise component can be identified as the Z source low frequency noise
(LFN) and very-low frequency noise (VLFN). Their respective
parameters and their relative contribution to the power spectrum
changed in ways comparable to  the other Z sources \citep[see, e.g.,
Fig.\ 5 in ][]{hovajo2002}. Although the LFN was slightly stronger
than in the other Z sources, this is mainly the result of the
slightly higher energy band we used in our analysis  (e.g.\ for PDS
I--III the rms of the LFN in the 2--60 keV band  was a factor
$\sim$1.4 lower than in the 6.9--60 keV band) and to a lesser extent
due to a different lower frequency of the range in which we measured
the rms (0 Hz, instead of 0.1 or 1 Hz).

In addition to the LFN and VLFN we also detected  two types of
low-frequency QPOs, both of which are clearly visible in Figure
\ref{fig:pds}, and kHz QPOs (not seen in interval B). These QPOs are
described in more detail in the remainder of this section.

\subsubsection{Horizontal branch oscillations}

A variable QPO can be seen clearly in PDS I--VI. It increased in
frequency from $\sim$9.5 Hz on the HB upturn to $\sim$53 Hz on the
upper NB as its strength decreased from $\sim$11\% rms to $\sim$4\%
rms. Its Q-value remained close to 6 in PDS I--III after which it
decreased to a value of $\sim$1.7 in PDS VI. In PDS I--III there were
indications for broad features (Q$\sim$1.5--2.5) at twice the QPO
frequency, which were likely the QPO's second harmonic. The frequency
range of the main QPO, the presence of a second harmonic, and the
relation of its frequency to position in the track was very similar
to that of the HBO in the Z sources. The Q-values and rms amplitudes
were slightly higher than those of the  HBO in the other Z sources,
but like in the case of the LFN this was the result of our choice of
energy band. We therefore identify this QPO as the Z source HBO.

HBOs were also detected on the HB of other intervals. However, they
were most prominent on the HB of the Cyg-like intervals. In none of
the other intervals the HBO was observed at frequencies as low as in
interval B, with interval C being the closest ($\nu_{max}\sim$11.5
Hz). The highest observed frequencies were near 59 Hz, in interval H.

\subsubsection{Normal branch oscillations}

\begin{figure}[t]
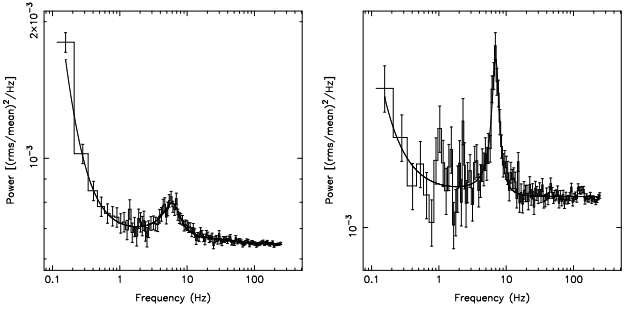
 \centerline{\hbox{
\includegraphics[width=4.0cm]{f7a.eps} \hspace{0.1cm}
\includegraphics[width=4.0cm]{f7b.eps}}} \caption{Two examples of
normal branch QPOs, from intervals B (left) and E (right). The solid
line in each panel shows the best fit to the power spectrum. The
Poisson levels have not been subtracted.}\label{fig:nbo} \end{figure}

Another, broad (Q$\sim$1.2) feature was detected at 6.5$\pm$0.3 Hz on the
middle/lower NB (PDS VII, see also Figure \ref{fig:nbo}).  Depending
on the type of interval, broad or narrow features around 7 Hz were
detected on the lower NB of almost every interval, except for
intervals F (which had poor coverage of the lower NB) and J. In all
cases these features were more prominent in the 6.9--60 keV band than
in the 2--60 keV band. 

The Cyg-like intervals A--D generally showed features with Q-values
less than 1.5, with interval C showing the broadest feature
(Q$\sim$0.9). Frequencies were around 6.3 Hz (A--C) or 8.1 Hz (D),
with fractional rms amplitudes of 3.0--3.5\%. An example from
interval B is shown in Figure \ref{fig:nbo}. The `intermediate'
intervals E, H, and I showed features with a higher coherence
(Q$\sim$1.8--3.2) that looked similar to the NBOs seen in, e.g., GX
340+0, GX 17+2, and Sco X-1 (see Fig.\ \ref{fig:nbo} for an example
from interval E). The frequencies of these narrower QPOs ranged
between 6.9 Hz and 7.3 Hz, with fractional amplitudes between 2.2\%
and 3.6\% rms. Based on the frequency range, Q-values, and location
on Z track, we identify these narrow QPOs, as well as the broader
features around 7 Hz in the Cyg-like intervals, as the Z source
normal branch oscillations (NBO).  The Sco-like intervals (F,G,J)
provided only one NBO detection (in interval G) at 9.0$\pm$0.7 Hz,
with a Q-value of 2.5$\pm$1.1 and an rms amplitude of 1.8$\pm$0.3\%.
Given the small number of detections in each interval, it was
difficult to study the dependence of the NBO properties on the
position along the Z track. 

Unlike what is seen in, e.g., GX 17+2 and Sco X-1, the NBOs in \sj\
did not evolve into flaring branch oscillations. No indications of
such QPOs were found on the flaring branches of any of the intervals.

\subsubsection{kHz QPOs}

\begin{figure}[t]
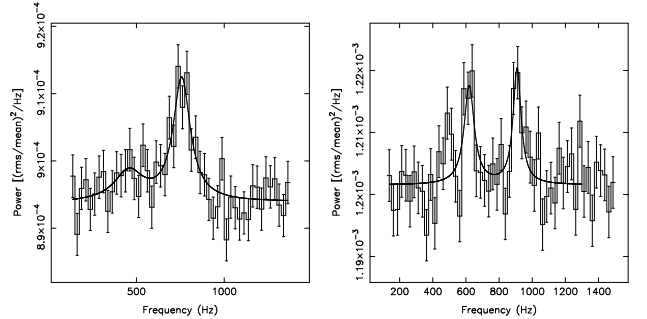
 \centerline{\hbox{
\includegraphics[width=4.0cm]{f8a.eps} \hspace{0.1cm}
\includegraphics[width=4.0cm]{f8b.eps}}} \caption{The two most
significant detections of kHz QPOs in \sj\ from observations
92405-01-03-05 on March 7 (left) and 92405-01-03-05 on March 21
(right). The solid line in each panel shows the best fit to the power
spectrum. The Poisson levels have not been subtracted. The small
excess around 480 Hz in the right panel is not statistically
significant.}\label{fig:khz} \end{figure}

We visually inspected all 2--60 keV and 6.9--60 keV power spectra for
(pairs of) kHz QPOs, like those seen in the atoll and Z sources
\citep{va2006}.  In addition to the (marginally significant) pair of
kHz QPOs reported by \citet{hobeva2006b} we found indications for
three additional pairs and made one very significant (6.1$\sigma$)
detection of a single peak.  The single peak (see Fig.\
\ref{fig:khz}) was detected in observation 92405-01-03-05 (March 7)
at 760$\pm$8 Hz, with a Q-value of 6.3$\pm$1.2 and a fractional
amplitude of 5.8$\pm$0.5\% rms (6.9--60 keV). An additional broad
peak (Q$\sim$2.3) was possibly present (2.7$\sigma$) at 470$\pm$50
Hz. Other twin kHz QPOs were only marginally detected, with single
trial significances for the individual peaks ranging from only
2$\sigma$ to 4$\sigma$. The most significant pair (4$\sigma$ and
3.8$\sigma$, see Fig.\ \ref{fig:khz}) was detected in observation
92405-01-03-05 (March 21); the frequencies were 620$\pm$10 Hz and
909$\pm$8 Hz, with Q-values of 8$\pm$3 and 16$\pm$6, and rms
amplitudes of 4.2$\pm$0.6\% and 4.0$\pm$0.6\%. The peak separation of
this pair was 288$\pm$13 Hz; the other (marginally detected) pairs
gave values of 293$\pm$56 Hz, 291$\pm$25 Hz, 249$\pm$18 Hz, and
265$\pm$35Hz. No significant dependence of peak separation on kHz QPO
frequency was found. We note that all the pairs of kHz QPOs we have
detected (without reference to expected frequency-frequency
relations), had frequencies matching the frequency-frequency
relations of other Z, and atoll, sources \citep[see,
e.g.,][]{psbeva1999}.

All (marginal) detections of kHz QPOs were made in observations on
the HB and only in intervals that were Sco-like or intermediate
between Sco-like and Cyg-like. For each of our intervals we  tried to
increase our sensitivity to kHz QPO by combining HB data from
multiple observations (selections were made in the SID). This did not
result in additional detections. We determined upper limits (95\%
confidence) on the strength of possible kHz QPO in the HB power
spectra (PDS I--IV) of interval B in the 100--1500 Hz range, with Q
values fixed to 2, 5, and 8. We found rms values ranging from 1.7\%
to 6.3\%.

\section{Discussion} 

During the first ten weeks of our {\it RXTE}/PCA observations of \sj,
the source showed all the characteristic X-ray color and variability
properties commonly associated with the  Z sources. In the CD
three-branched tracks were traced out, which, although variable in
shape, often resembled those of the Z sources (\S
\ref{sec:results_curves}), and the power spectra showed various noise
components and QPOs that were similar, in terms of properties and
evolution along the Z track, to those seen in the Z sources (i.e.
LFN, VLFN, HBO, NBO, and kHz QPOs; \S \ref{sec:results_variability}).

\subsection{Comparison with other Z sources}\label{sec:compare}

\subsubsection{Color-color diagrams}

\sj\ displayed a surprisingly large variety in the shape of its Z
tracks in the CD. During the first part of the outburst (intervals
A--D) the Z tracks showed similarities to those of the Cyg-like Z
sources, in particular GX 5-1 \citep{jovaho2002} and GX 340+0
\citep{jovawi2000}. These similarities included an upturn on the HB
and a downturn on the FB, which has previously been referred to as
the `dipping' FB \citep[no clear dips were seen in the light
curves of \sj]{kuva1995,jovaho2002}. In the second part of our data set the Z
tracks were more similar to those of the Sco-like Z sources
(intervals F, G, and J) with the FB being more prominent than in the
first intervals.

\subsubsection{Hardness-intensity diagrams}

Comparing the SID/HID tracks of \sj\ with those of the Z sources
revealed some minor differences. For example, the HID tracks of the
Cyg-like intervals appear a bit different from those of GX 5-1 and GX
340+0; in \sj\ the count rate decreased from the HB/NB vertex to the
end of the FB, whereas GX 5-1 and GX 340+0 showed a decrease from the
HB/NB vertex to the NB/FB vertex, followed by an increase to the
middle of the FB and dipping at the top of the FB. Cyg X-2, on the
other hand, does occasionally show similar count rate behavior as
\sj\ did in its Cyg-like intervals \citep{wiva2001}. Although changes
near the middle of the FB were also seen in \sj\ (interval F) the
observed behavior was the opposite of GX 5-1 and GX 340+0 (decrease
in count rate on the FB followed by an increase). Among the Sco-like
intervals we observed some variety in the HID tracks. Intervals G and
J were similar to the Sco-like Z sources, but interval F seemed to
combine some properties of the Sco-like and Cyg-like sources. All
these differences show that similarities in the shapes of Z tracks in
the CD do not necessarily extend fully to the HID and SID.

\subsubsection{Evolution of the Z tracks}

The NB/FB vertex was the only vertex observed in all intervals and
can serve as a reference point for following the evolution of the Z
track. We estimated the count rate at the NB/FB vertex for each of
the intervals, which led to the following ordering (from high to low
count rates): C, B/D, A, E, F/H, G/I, and J. Following this sequence
we observed that the Z tracks shifted to harder colors as the source
became weaker (see Fig.\ \ref{fig:ccd}), which has also been seen in
Cyg X-2 \citep{wivaku1997}. Another, very clear change occurred in
the orientation of FB in the CD, pointing downwards when the source
was brightest and rotating counter-clockwise as the source became
weaker. In the SID the FB turned clockwise from the left (C) to upper
left (B/D, A), then partially overlapped with the NB (E, F/H) and
finally pointed to upper right (G, J). At the same time the range in
count rates observed on the HB and NB became much smaller and the HB
changed from pointing to the upper left to a nearly vertical position
in the HID.

Changes in the shape and/or position of the tracks in the CDs, SIDs,
and HIDs (secular changes) have been observed in GX 17+2 and the
Cyg-like sources, with Cyg X-2 showing the most dramatic changes
\citep{kuvava1996,wivaku1997}.  In Cyg X-2 it was found that the FB
changed from pointing to the left in the HID to pointing to the upper
left when the source changed between its high and medium intensity
levels, i.e., in the same sense relative to flux level as we observe
in the SID of \sj. On one occasion \citet{kuvava1996} observed a
flaring-like branch, with a shape similar to that of the FB observed in
the Sco-like source. This occurred at the lowest intensity levels,
also in line with what we observed for \sj.

\subsubsection{Variability}

A detailed timing analysis of all intervals needs to be done to
compare the properties of individual variability components, but at
first sight the evolution of the variability properties did not seem
to be affected strongly by the changes in Z track itself. There were
however changes in the presence of certain types of variability.
First, there was a clear evolution in the properties of the NBO,
changing from a broad feature in the Cyg-like intervals to a narrow
feature in the intermediate intervals and perhaps disappearing in the
Sco-like intervals. Second, kHz QPOs were only detected in the
intermediate and Sco-like intervals, despite similar or even higher
count rates in the Cyg-like intervals and the fact that kHz QPOs have
been found in all three Cyg-like Z sources
\citep{wihova1998,jowiva1998,wimeva1998}. However, it is interesting
to note that the kHz QPOs of Cyg X-2 where discovered when the source
showed CD tracks similar to those of the Sco-like intervals F and G
\citep{wihova1999}, but they were not found when its CD tracks were
more similar to those of the Cyg-like intervals B and C
\citep{wiva2001}.

\subsection{The role of \mdot\ in secular changes and Z/atoll
types}\label{sec:disc_mdot}

As discussed earlier, \sj\ is not the first Z source to show large
secular changes, but it is the first in which these changes resulted
in a switch between full-blown Cyg-like and Sco-like behavior; Cyg
X-2 has also shown a Sco-like flaring branch
\citep{kuvava1996,murech2001}, but not a full Sco-like Z track.
Observing both types of behavior in a single source puts a different
perspective on two previously proposed explanations for the existence
of two types of Z sources and also on the origin of the secular
changes.

\citet{kuvaoo1994} suggested that the difference between the Cyg-like
and Sco-like Z sources is the result of a lower binary inclination in
the latter group. Our observations of \sj\ would require a change in
the binary inclination on a time scale of a few weeks, much faster
than expected (and requiring interaction with a third stellar
component). The accretion disk itself might be precessing on such a
time scale, possibly leading to different viewing angles of the inner
accretion flow \citep[as originally suggested by][for Cyg
X-2]{vrswke1988}. However, this would likely only have an observable
effect in sources with a high binary inclination, for which we have
seen no indications (such as dips or eclipses). Moreover, at some
point, when the inner disk returns to its original inclination, \sj\
would have to switch back to Cyg-like behavior, which has not yet
been seen at the time of writing. \citet{pslami1995} explained the
difference between the two groups in terms of the  neutron star
magnetic field, which should be smaller in the Sco-like group. Again,
our observations of \sj\ require changes on a time scale of a few
weeks, much faster than is expected for magnetic field decay.
Another, somewhat related option might be magnetic screening, which
causes the magnetic field  to become buried by the freshly accreted
material at high \mdot\ \citep[see, e.g.,][]{cuzwbi2001}. However, at
near-Eddington \mdot\ the effects of the screening are much larger
(orders of magnitude decrease) than the expected difference (if any
exists at all) between the Cyg-like and Sco-like sources. Moreover,
the fact that we see two types of persistent Z sources, accreting at
near-Eddington rates, also rules magnetic screening out as a
explanation, because the magnetic field should be buried in both
types of sources.

Since the above explanations do not work, we have to search for
another parameter that might result in two types of Z sources as well
as the observed switch in \sj. Given the transient nature of \sj\
\mdot\ seems an obvious candidate, not only to explain the
differences among the Z sources, but also as the origin of the
secular changes seen in individual sources \citep[as suggested
by][]{hovajo2002}. This option has long been disregarded as a viable
alternative because it was generally assumed that \mdot\ is already
responsible for changes along the Z track. However, the possibility
of \mdot\ causing both changes along the Z track {\it and} changes in
the shape of the Z track itself is not necessarily a problematic one.
As a solution to the problem of the `parallel tracks' in NSXBs
\citep[e.g.][]{mevava1998a}, \citet{va2001} suggested a model in 
which there exists both a prompt and a filtered response to changes
in the \mdot\ through the disk, \mdisk. Suggested mechanisms
responsible for such a filtered response are a radial inflow or
nuclear burning. In the cases of the `parallel tracks', the kHz QPO
frequency is then determined by the ratio of the \mdisk\ and filtered
\mdisk, \mfilt,  and the luminosity is determined by \mdisk\ plus a
contribution from \mfilt. Following \citet{va2001}, we apply this
model to \sj\ and suggest that the ratio \mdisk/\mfilt determines
the position along the Z (which in Z sources also means the kHz QPO
frequency). Motion along the Z is then the result of spectral changes
caused by variations in \mdisk/\mfilt. The details of how these
spectral changes occur, which determine the shape of the Z tracks,
depend on the underlying physical components, whose properties vary
with \mdisk, \mfilt, or a combination of the two. Hence, as the
luminosity changes (due to changes in \mdisk\ and/or \mfilt), the
shape of the Z tracks is likely to change, as we observe \sj.

As long as the total \mdot\ does not change too much, the overall
shape of the Z is expected to remain similar. Assuming that the count
rates at the NB/FB vertex in each interval are representative for the
\mdot\ during those periods, we attempt to estimate the magnitude of
the changes in \mdot\ that are required to change the appearance of
the Z track. Interval C probably had the highest \mdot\ (see
\S\ref{sec:compare}), so Cyg-like behavior can probably be associated
with the highest \mdot\ range. The count rates at the NB/FB vertex in
the Sco-like interval G were about a factor of 1.5 lower than those
in the C,  suggesting that changes in the count rate (and perhaps
\mdot) by factors of 1.5 or less might already be enough to switch
between the two types of Z source behavior. Taking the peak count
rates of these two intervals, instead of the NB/FB count rates, gave
a similar factor (1.7).

In the past the luminosities at the vertices of the Z have been
treated as standard candles (possibly related to \ledd), to estimate
the distances of the Z sources \citep[see, e.g.,][]{brgefo2003}. Our
\sj\ observations show that probably none of these vertices can be
uniquely identified with \ledd\ or any other constant luminosity. In
fact, they suggest that {\it the Cyg-like Z sources might be
substantially more luminous ($>$50\%) then the Sco-like Z sources},
which affects distance estimates that are based on the assumption
that all Z sources have similar luminosities. Since Sco X-1, for
which an accurate distance has been established through parallax
measurements \citep{brfoge1999}, is known to have super-Eddington
luminosities occasionally \citep[for a 1.4 $M_\odot$;][]{brgefo2003},
the Cyg-like Z sources are probably even further above the Eddington
limit.

\begin{figure}[t] 
\centerline{\includegraphics[width=8.0cm]{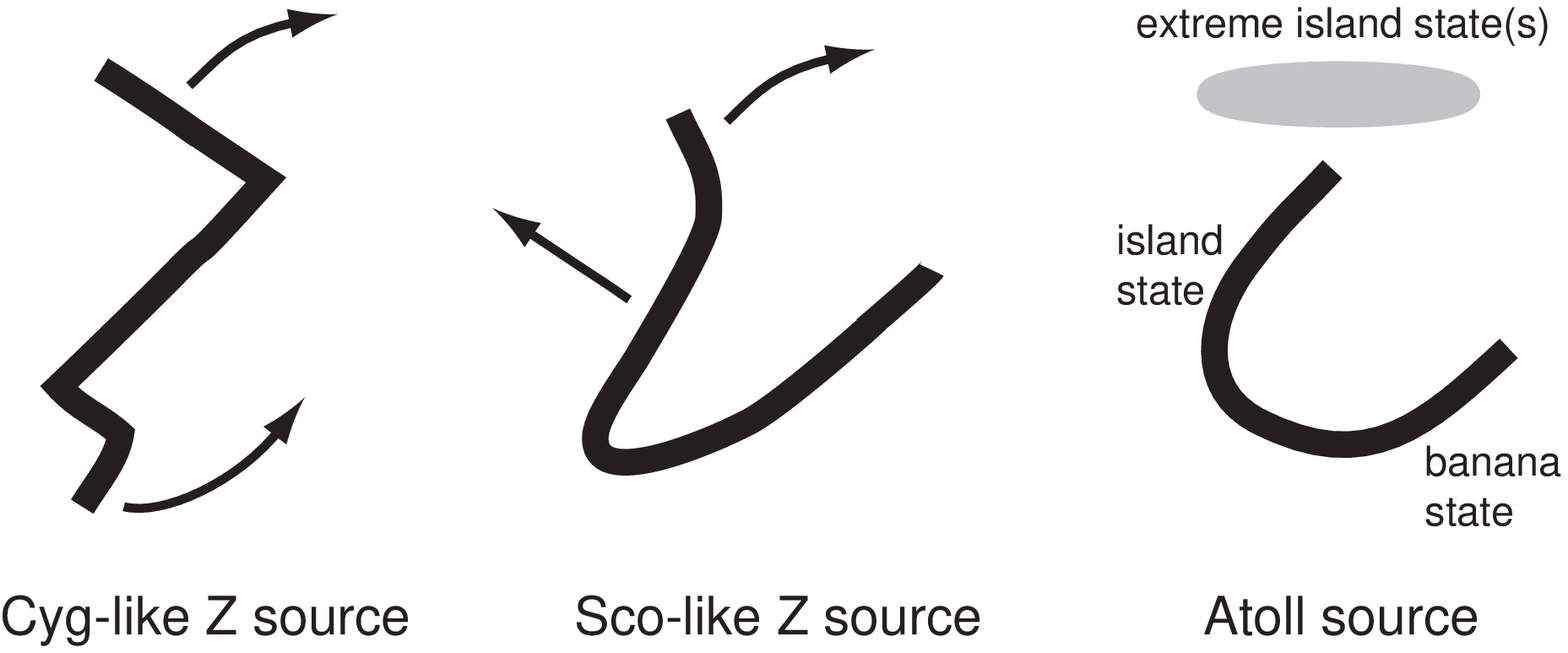}}
\caption{Possible evolution of the CD tracks of a neutron star LMXB
as a function of decreasing \mdot. The arrows indicate how the
different branches of the (black) tracks changes their orientation.
The grey area at the top of the atoll track represent the extreme
island state that is only seen a few atoll sources and which was
interpreted by \citet{murech2001} and \citet{gido2002} as the
equivalent of the Z source Horizontal Branch. An analysis by 
\citet{vavame2003} has shown that the island state can connect to
different locations along the extreme island
state.}\label{fig:evolution} \end{figure}

It is likely that \sj\ eventually decreases to count rates well below
the range observed for the Sco-like intervals and at that point it
may turn into an atoll source. Based on the observed change between
the two types of Z behavior this will probably be a gradual process
as well. \sj\ has shown that the orientation and shape of the
branches in the X-ray color diagrams change as a function of the
overall count rate level. Extrapolating this behavior to lower count
rates suggests that the branches of the Z track could slowly
transform into the branches seen in atoll sources.  In Figure
\ref{fig:evolution} we have drawn a simple cartoon that depicts how
the tracks in a CD might evolve as a function of decreasing \mdot.
The HB of the Cyg-like Z tracks rotates clockwise from the upper left
to the upper right and together with the NB it transforms into the
atoll island state. The FB transforms from a short downward pointing
branch into a long upward curved branch that becomes the atoll banana
state. Of course, this proposed evolution is purely based on the
shape of the tracks in the CDs. A detailed timing analysis is still
necessary to check these claims. Moreover, sofar we have only
described the observed spectral evolution in terms of
phenomenological changes (i.e. changing position and orientation of
the different branches). Spectral fits are necessary to relate these
phenomenological changes to physical changes in, e.g., the boundary
layer or Comptonizing medium.

As \mdot\ decreases,  \sj\ may first go through a phase in which it
resembles the bright atoll sources GX 9+1, GX 9+9, GX 3+1, and in
particular GX 13+1, which appears to be a hybrid atoll/Z source
\citep{screva2003,howiru2004}, and then possibly starts showing the
characteristic atoll island/banana states \citep{hava1989}. Note
that, although the CD track of interval J looks similar to the ones
traced out by GX 13+1 \citep{screva2003}, the timing properties in
this interval are different from GX 13+1. If ideas attributing Z
behavior to a relatively stronger B field are correct then at lower
luminosity, where scattering in the accretion flow preventing us from
detecting the pulsations would be less, a (possibly millisecond)
pulsar might be observed.  It is also quite possible that X-ray
bursts will occur at some point. Burst and pulsations have not been
seen in \sj\ at the time of writing.

Based on the (nearly) horizontal orientation of the extreme island
state in the CDs of some atoll sources, \citet{murech2001} and
\citet{gido2002} identified this state with the Z source HB (see also
the cartoon in Fig.\ \ref{fig:evolution}). In addition to the
concerns raised by other authors
\citep{baol2002,vavame2003,revava2004,va2006}, we note that at lower
\mdot\ (in the Sco-like intervals) the HB no longer seems to be
horizontal. So, unless the HB would return to its original horizontal
position at even lower \mdot, for which we see no evidence in the
more recent data of \sj, the extreme island state seen in some atolls
is probably not related to the HB seen in the Z sources.  Perhaps the
HB upturn observed in intervals B and C might related to the extreme
island state, but this upturn was not observed often enough to test
this. 

\subsection{The nature of \sj}

Why is \sj\ the only transient Z source thus far? The persistent Z
sources differ from the persistent atoll sources in that they have a
much higher \mdot. It has been suggested that this could be the
result of differences in the properties of the secondary star. The
long orbital periods found for Sco X-1 \citep[18.9 hr,][]{gowrli1975}
and Cyg X-2 \citep[9.84 days,][]{cocrhu1979} suggest that the
companion stars in Z sources are evolved stars, resulting in higher
mass transfer rates. What matters for the properties of transient
outbursts (e.g. duration and perhaps peak luminosity) is not the mass
transfer rate from the secondary, but how much matter has been stored
in the accretion disk before an outburst starts. Obviously, systems
with longer orbital periods result in larger Roche lobes for the
neutron star, which allows for larger accretion disks that presumably
can store more matter. Calculating the exact properties of such a
system is beyond the scope of the current paper. However, given the
fact that at the time of writing \sj\ has been accreting at near
Eddington levels for more than 215 days, i.e. much brighter and for a
longer period than most transient NSXBs, it is likely that, just like
some of its persistent counterparts, the transient Z source \sj\ has
a long orbital period. At this point no clear signatures of an
orbital period have been found in X-ray or optical/IR light curves.
The best indications for possible modulations were observed in the
{\it RXTE}/ASM (see Fig.\ \ref{fig:asm}) and {\it RXTE}/PCA light
curves between MJD $\sim$53820 and MJD $\sim$53920. These modulations
had a (quasi-)period of $\sim$25 days, but they are no longer present
at the time of writing and their nature is unclear. We probably have
to wait until the source has reached quiescence to search for optical
signatures of an orbital period. In any case, the fact that \sj\ is
the first transient Z source would suggest that the number of sources
with long orbital periods is small, in line with small number of the
persistent Z sources.

Assuming that the shape of Z tracks changes with luminosity in the
same way in all Z sources we can attempt to estimate the distance of
\sj. We do this by comparing the flux of \sj\ to that of Sco X-1,
which is the only Z source with a well determined distance
\citep[2.8$\pm$0.3 kpc, as determined by radio parallax
measurements;][]{brfoge1999}. For our estimate we use the track of
interval G, which, despite a small difference in the orientation of
the HB, resembled the track of Sco X-1 most \citep{brgefo2003}.  We
choose the NB/FB vertex as a reference point. For Sco X-1 we measured
a 3--25 keV keV luminosity of $2.6\times10^{38}$ erg\,s$^{-1}$ at
that vertex (obs-ID: 40706-02-06-00); we find an unabsorbed 3--25 keV
flux for \sj\ of $1.01\times10^{-8}$ erg\,s$^{-1}$\,cm$^{-2}$
(obs-ID: 92405-01-01-03), which implies a source distance of 14.7
kpc, with an uncertainty of at least 20\%. For that distance, the
highest 2--18 keV luminosity measured in our data set (i.e. near the
HB/NB vertex in interval C, obs-ID: 91106-02-03-14) would be
$\sim$$7.1\times10^{38}$ erg\,s$^{-1}$, well above the Eddington
limit of a 1.4 $M_\odot$ neutron star. Note, again, that both the
distance and luminosity estimates rely on assumptions that cannot be
verified easily.

\section{Summary \& Conclusions}

We observed \sj\ during the early phase of its first known outburst.
Based on its correlated X-ray color and timing properties we
concluded that during this period the source could be classified as a
Z source, making it the first transient neutron star X-ray binary
with such properties. No X-ray bursts or pulsations have been
observed at the time of writing.

A clear change was observed in the tracks traced out in X-ray color
diagrams. Initially these looked very similar to those of the
Cyg-like Z sources (GX 5-1, GX 340+0, and Cyg X-2), but later they
gradually transformed into tracks similar to those seen in the
Sco-like Z sources (GX 17+2, GX 349+2, and Sco X-1). We observed
noise components and QPOs such as commonly also seen in other Z
sources and found that the properties and/or presence of the NBO and
kHz QPOs depended on the shape of the Z track. However, the general
evolution of the variability properties along the Z track did not
change much between the different tracks.

The transient nature of \sj\ and the associated changes in \mdot\ are
likely responsible for the large changes in the Z tracks, which we
interpret as extreme cases of the secular changes observed in the
persistent Z sources. This interpretation requires both a prompt and
a filtered  response to changes in the \mdot\ if one assumes that
\mdot\ also correlates with position along the Z track. As \sj\
continues its decay, the branches of the Z are likely to transform further.

Cyg-like behavior was observed at the highest count rates and
presumably the highest \mdot, suggesting that the Cyg-like Z sources
might be intrinsically more luminous than the Sco-like Z sources. 

\sj\ provides us with a unique opportunity to follow the behavior of
accretion flows onto neutron stars from (super-)Eddington accretion
rates all the way to quiescence. The first ten weeks of our {\it
RXTE} observations already allowed us to construct a sequence of Z
tracks based on overall count rate levels. Hopefully, as the source
decays, this sequence can be extended to include different types of
phenomenology such as perhaps atoll-like tracks, X-ray bursts or even
pulsations. Such a sequence would be of great importance for
understanding the nature of the different types of persistent NSXBs.

\acknowledgments

The authors would like to thank Jean Swank and Evan Smith for
scheduling the daily {\it RXTE} observations of \sj, Ron Remillard
for stimulating discussions, Erik Kuulkers and Peter Jonker for
their feedback on an earlier version of this paper, and, finally, the
referee for carefully reading our manuscript and providing helpful
comments. This research has made use of data obtained from the High
Energy Astrophysics Science Archive Research Center (HEASARC),
provided by NASA's Goddard Space Flight Center. E.G is supported by
NASA through {\it Chandra} Postdoctoral Fellowship Award PF5-60037,
issued by the {\it Chandra} X-Ray Observatory Center, which is
operated by the Smithsonian Astrophysical Observatory for and on
behalf of NASA under contract NAS8-39073.


\end{document}